\documentclass[cameraready]{Interspeech}
\usepackage{multirow}   
\usepackage{array}      
\usepackage{tabularx}
\usepackage{makecell}
\usepackage{url}

\title{Schr\"odinger Bridge Mamba for One-Step Speech Enhancement}

\author[orcid=0000-0002-2565-0574]{Jing}{Yang} 
\author[orcid=0009-0009-9660-4610]{Sirui}{Wang}
\author[orcid=0009-0008-1771-935X]{Chao}{Wu}
\author[orcid=0000-0001-5134-4506]{Lei}{Guo}
\author[orcid=0009-0008-4407-2483]{Fan}{Fan}

\address{Central Media Technology Institute, Huawei}

\email{(yangjing201, wangsirui1, wuchao17, guolei94, fanfan1)@huawei.com}

\keywords{Schrödinger Bridge, Mamba, state-space models, trajectory modeling, speech enhancement}

\usepackage{comment}
\newcommand{\tablescriptsize}{\fontsize{6pt}{7.2pt}\selectfont}

\begin{document}
\maketitle

\begin{abstract}
We present Schrödinger Bridge Mamba (SBM), a novel model for efficient speech enhancement by integrating the Schrödinger Bridge (SB) training paradigm and the Mamba architecture. Experiments of joint denoising and dereverberation tasks demonstrate SBM outperforms strong generative and discriminative methods on multiple metrics with only one step of inference while achieving a competitive real-time factor for streaming feasibility. Ablation studies reveal that the SB paradigm consistently yields improved performance across diverse architectures over conventional mapping. Furthermore, Mamba exhibits a stronger performance under the SB paradigm compared to Multi-Head Self-Attention (MHSA) and Long Short-Term Memory (LSTM) backbones. These findings highlight the synergy between the Mamba architecture and the SB trajectory-based training, providing a high-quality solution for real-world speech enhancement. 
Demo page: \url{https://sbmse.github.io} 
\end{abstract}

\section{Introduction}
\label{sec:intro}

\begin{figure*}[t!]
	\centering 
	\includegraphics[width=\textwidth]{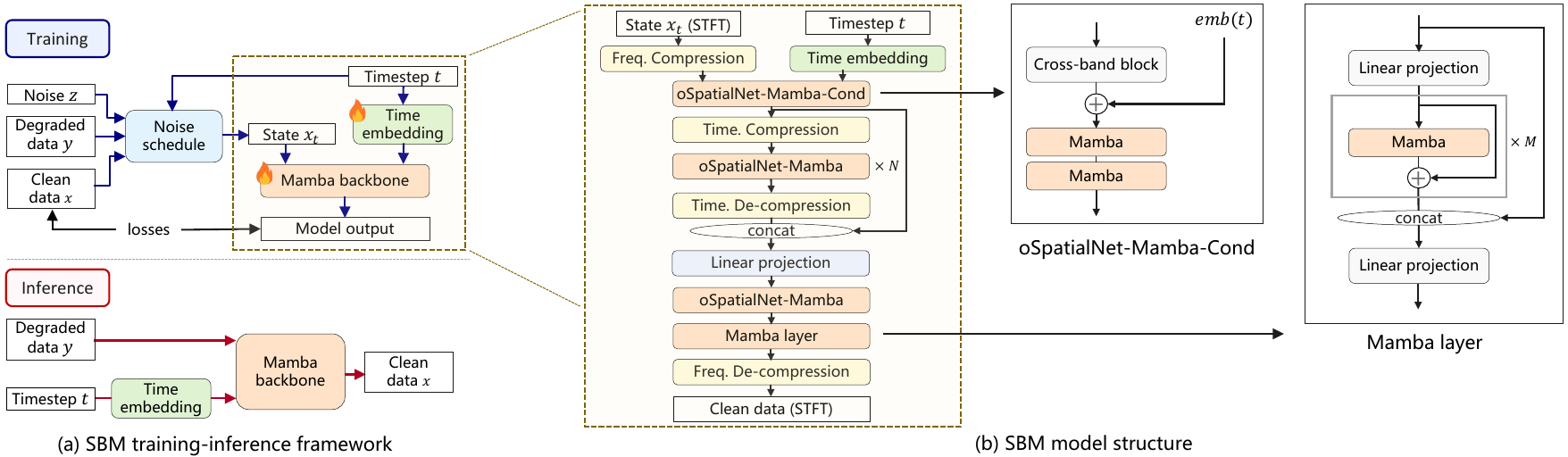}
	\caption{Overview of the SBM training-inference framework and the Mamba-based model architecture.}
	\label{fig:overview}
\end{figure*}

Deep generative models are increasingly employed for speech enhancement (SE). By learning the clean speech distribution directly, generative models achieve superior perceptual quality and reconstruct fine details lost in deterministic regression~\cite{richter2023speech}. 
While earlier score-based generative models showed promising SE performance, they suffer from the \textit{mean prior mismatch} issue~\cite{richter2023speech}. 
To resolve this, the Schrödinger Bridge (SB) paradigm models the optimal transport (OT) path via stochastic differential equations (SDEs), seamlessly transporting the degraded prior directly to the clean target~\cite{jukic2024schr,wang2024diffusion,richter2025investigating,han2025few,nishigori2025schrodinger}. 
SB methods have demonstrated remarkable success across SE tasks (denoising/dereverberation~\cite{jukic2024schr,wang2024diffusion,richter2025investigating,han2025few,nishigori2025schrodinger}, super-resolution~\cite{li2025bridge}, inpainting~\cite{kong2025a2sb}), image generation~\cite{liu20232} and text-to-speech~\cite{chen2023schrodinger}. 
Following broader successes~\cite{songscore,vahdat2021score}, most SB-based SE methods adopt the NCSN++ architecture as their backbone (SB-NCSN++)~\cite{jukic2024schr,richter2025investigating,han2025few,nishigori2025schrodinger}.

However, the slow inference of these SB-based methods (often requiring $>10$ iterative steps) hinders real-time application. To accelerate generation, recent approaches explored specialized schedulers and solvers~\cite{li2025bridge}, adversarial training~\cite{han2025few,xu2024ufogen} or consistency trajectory modeling~\cite{nishigori2025schrodinger,kim2023consistency}. While some achieve one-step inference in denoising or dereverberation tasks~\cite{han2025few,nishigori2025schrodinger}, existing works typically overlook the inherent synergy between the SB paradigm and the backbone architecture, leaving significant room for improvement in efficient modeling.

Recently, the selective state-space model Mamba~\cite{gu2023mamba} has emerged as a powerful architecture for modeling long-range audio dependencies. Although adopted in SE tasks (e.g., oSpatialNet~\cite{quan2024multichannel}, SEMamba~\cite{chao2024investigation}, USEMamba~\cite{chao2025universal}), these works utilize deterministic mapping or masking training strategy, failing to exploit the potential of generative trajectory learning. 

In this work, we propose \textbf{\textit{Schr\"odinger Bridge Mamba (SBM)}}, the first framework to synergize the Schrödinger Bridge (SB) paradigm with the Mamba architecture for one-step speech enhancement leveraging generative trajectory guidance. 
Our core argument is that aligning \textit{training paradigm} with the \textit{backbone architecture}'s inductive bias is essential for efficiency and effectiveness.
Unlike `blind' deterministic mapping or masking that only sees the start and end points, the SB paradigm explicitly calculates intermediate states $x_t$ along the optimal transport path bridging the degraded and clean speech distributions~\cite{jukic2024schr}. These states $x_t$ are utilized as training inputs, effectively acting as `anchors' to guide the backbone model's learning of the underlying evolution process. 
The Mamba architecture is well-suited for the SB training paradigm: (1) As a state-space model, Mamba naturally emulates the process of state evolution~\cite{gu2023mamba}; (2) Its selective mechanism facilitates adaptive context modeling, allowing the network to learn the dynamics of the optimal transport path~\cite{gu2023mamba,hu2024zigma}.
Overall, SBM `distills' the SB transformation into the Mamba dynamics during the training process, enabling high-fidelity reconstruction in a single inference step.

We evaluate SBM on joint denoising and dereverberation SE tasks that represent common real-life scenarios.
Experiments on DNS and VoiceBank-Demand test sets demonstrate that SBM outperforms conventional SB-NCSN++ models, mapping-trained Mamba model, flow-matching variant, one-step SB variants and strong discriminative model (ZipEnhancer~\cite{wang2025zipenhancer}) in real-world and reverberant scenarios, and achieves the best real-time factor (RTF) among these methods.
Furthermore, ablation studies replacing Mamba with Multi-Head Self-Attention (MHSA) and Long Short-Term Memory (LSTM) reveal that the SB paradigm consistently outperforms the mapping paradigm across architectures, highlighting the potential of the training paradigm. Moreover, the superior performance of Mamba in these experiments highlights its effectiveness in capturing trajectory dynamics.
Collectively, these results justify the SBM design and underscore the potential of integrating continuous-time diffusion processes with state-space backbones, offering valuable insights for the broader evolution of continuous-time sequence modeling in complex audio tasks.  

\section{Schr\"odinger Bridge Mamba}
\label{sec:method}
Schrödinger Bridge Mamba (SBM) leverages the generative optimal transport (OT) trajectory of Schrödinger Bridge (SB) to train a selective state-space Mamba backbone. This section details the SB formulation, the model architecture and the training-inference framework. 

\subsection{Schr\"odinger Bridge Formulation}
\label{sec:methodsub1}
Unlike standard diffusion models that rely on Gaussian priors leading to \textit{mean prior mismatch} issue~\cite{richter2023speech}, 
SBM formulates speech enhancement as an OT process directly between the degraded speech distribution $p_T$ and the clean speech distribution $p_0$. 
Mathematically, this process is governed by a system of SDEs, and the intermediate states $\{x_t\}_{0:T}$ along the SB stochastic process can be constructed using paired data samples (clean speech $x \sim p_0$ and degraded speech $y \sim p_T$)~\cite{jukic2024schr,schrodinger1932theorie,chen2021stochastic}. 
To enable efficient training for speech enhancement,
following the tractability of SB OT formulation~\cite{jukic2024schr, richter2025investigating,han2025few,nishigori2025schrodinger}, the intermediate states $x_t$ at timesteps $t \in [0,1]$ can be explicitly parameterized as an interpolation of the boundary conditions plus a stochastic Wiener process term. Specifically,
$x_t$ is sampled via 
\begin{equation}
x_t=\mu_x(t)+\sigma_x(t)z, \quad z \sim \mathcal{N}_C(0,\mathbf{I}) 
\end{equation}
where $\mu_x(t)=w_x(t)x+w_y(t)y$. 
The variance $\sigma_x(t)$ and coefficients $w_x(t)$, $w_y(t)$ are derived from the Variance Exploding (VE) noise schedule parameters~\cite{jukic2024schr} in our implementation.
Crucially, Eq.(1) serves as the data generator during training: For each batch, we sample random timesteps $t$ and compute states $x_t$, training the model to reconstruct the clean target $x$ by learning the underlying data distributions.

\subsection{Mamba-based Architecture}
\label{sec:methodsub2}
The architectural design of SBM is motivated by the structural resemblance between the SB theory and Mamba's discretized recurrence $h_t=\mathbf{A}h_{t-1}+\mathbf{B}u_t, y_t=\mathbf{C}h_t$~\cite{gu2023mamba}. 
Viewed as a controlled system, $\mathbf{A}h_{t-1}$ represents natural, uncontrolled dynamics~\cite{kalman1960new}, while the input-dependent parameters $(\mathbf{B}, \mathbf{C})$ act as control terms. Training Mamba thus resembles learning the optimal control strategy inherent to SB~\cite{chen2021stochastic}, where the selective mechanism dynamically parameterizes the transport path based on current states.
Furthermore, recently proven mathematical optimality of Mamba in capturing Markovian sequence dynamics~\cite{bondaschi2025markov} also suggests its suitability for modeling the stochastic evolution process of the SB paradigm~\cite{chen2021stochastic}. 

Guided by these insights, we construct a Mamba-based model as shown in Figure~\ref{fig:overview}(b).
Following the common practice in related work~\cite{jukic2024schr,richter2025investigating,han2025few,nishigori2025schrodinger,quan2024multichannel,chao2024investigation,chao2025universal}, we represent audio using STFT spectra. 
The main component \textit{oSpatialNet-Mamba} is implemented following the design in~\cite{quan2024multichannel}.
To incorporate the timesteps defined by the SB training paradigm, we train a \textit{Gaussian Fourier} module to embed timesteps~\cite{jukic2024schr,richter2025investigating}, and the time embedding \textit{emb(t)} is added to the input of Mamba in the \textit{oSpatialNet-Mamba} module, which is termed \textit{oSpatialNet-Mamba-Cond} in SBM.
Inspired by~\cite{quan2024multichannel,quan2024spatialnet}, to address the temporal limitations of oSpatialNet, we integrate a fullband \textit{Mamba layer} to capture global spectral dynamics and inter-frame dependencies.
For streaming feasibility, the Mamba backbone operates with a small lookahead of 2-4 frames, yielding an algorithmic latency under \qty{40}{ms} to balance causality and enhancement quality.


\subsection{Training and One-Step Inference}
\label{sec:methodsub3}
Figure~\ref{fig:overview}(a) sketches the training and inference process of SBM. The Mamba backbone takes noisy intermediate states $x_t$ and timestep embeddings as inputs to learn the distribution of clean data $x$. 
Typical denoising objectives as in diffusion models can be applied~\cite{han2025few}, among which the data prediction loss has shown good performance for speech enhancement~\cite{jukic2024schr,richter2025investigating,han2025few,nishigori2025schrodinger}. 
We employ a comprehensive data prediction loss combining magnitude and complex domain constraints: $L=\lambda_1L_{mse}(S,\hat{S})+\lambda_2L_{mse}(\lvert S \rvert,\lvert \hat{S} \rvert)+\lambda_3L_{mr,mse}(S,\hat{S})+\lambda_4L_{mr,mse}(\lvert S \rvert,\vert \hat{S} \rvert)$, where $S$ and $\hat{S}$ are the target and modeled STFT, $mr$ refers to multi-resolution and $\lvert S \rvert$ refers to magnitude spectrum.

While standard SB methods require iterative solving of the reverse SDE, SBM is designed for one-step generation. During inference, we set the timestep to the start of the reverse process ($t=1$, corresponding to the degraded prior). The model directly reconstructs the clean target in a single forward pass. 
This drastically reduces latency while leveraging the distributional modeling capability learned from the SB trajectory.

\begin{table*}[t]
	\centering
	\caption{Performance comparison on benchmark test sets DNS With Reverb, DNS No Reverb and VoiceBank-Demand. ``RTF" stands for real-time factor, measured on a GPU by averaging over 10 pieces of 10s audio samples.}
    \tablescriptsize
	\renewcommand\arraystretch{0.75}
	\begin{tabular}{p{1.3cm} l c c c c c c c c c c c}
		\toprule
		Test set & Model & Para.[M] & RTF$\downarrow$ & SIG$\uparrow$ & BAK$\uparrow$ & OVRL$\uparrow$ & P808MOS$\uparrow$ & NISQA$\uparrow$ & SpeechBERTScore$\uparrow$ & Similarity$\uparrow$ & PESQ$\uparrow$ & ESTOI$\uparrow$ \\
		\midrule
		\multirow{9}{1.3cm}{DNS With Reverb}
		& SB-NCSN++(50) & 25.16 & 0.767 & 3.395 & 3.938 & 3.085 & 3.884 & 3.879 & 0.703 & 0.947 & 1.532 & 0.629 \\
		& SB-NCSN++(10) & 25.16 & 0.156 & 3.425 & 3.97 & 3.126 & 3.896 & \textbf{\underline{4.087}} & 0.728 & 0.947 & 1.597 & 0.671 \\
		& SB-NCSN++(1) & 25.16 & 0.0155 & 3.281 & 3.871 & 2.948 & 3.559 & 3.323 & 0.730 & 0.943 & 1.586 & 0.67 \\
		& SBCTM & 65.98 & 0.02 & 2.927 & 3.859 & 2.623 & 3.035 & 2.578 & 0.494 & 0.859 & 1.218 & 0.281 \\
		& SB-UFOGen & 25.16 & 0.0152 & 2.964 & 3.664 & 2.596 & 3.337 & 2.131 & 0.662 & 0.918 & 1.383 & 0.601 \\
        & ZipEnhancer & 2.04 & 0.105 & 2.771 & 3.164 & 2.264 & 2.85 & 1.926 & 0.534 & 0.84 & 1.227 & 0.288 \\
		& Mamba-base & 3.61 & \textbf{\underline{0.0048}} & 3.202 & 3.898 & 2.895 & 3.665 & 3.514 & 0.766 & 0.963 & 1.895 & 0.696 \\
        & FM-Mamba & 3.93 & \textbf{\underline{0.0048}} & 3.164 & 3.956 & 2.887 & 3.461 & 3.148 & 0.682 & 0.930 & 1.643 & 0.634 \\
		& SBM & 3.93 & \textbf{\underline{0.0048}} & \textbf{\underline{3.653}} & \textbf{\underline{3.995}} & \textbf{\underline{3.367}} & \textbf{\underline{4.055}} & 4.053 & \textbf{\underline{0.784}} & \textbf{\underline{0.965}} & \textbf{\underline{1.971}} & \textbf{\underline{0.71}} \\
		\midrule		
		\multirow{9}{1.3cm}{DNS No Reverb}
		& SB-NCSN++(50) & 25.16 & 0.767 & 3.515 & 4.02 & 3.231 & 4.01 & 4.233 & 0.78 & 0.976 & 1.656 & 0.8 \\
		& SB-NCSN++(10) & 25.16 & 0.156 & 3.477 & 4.069 & 3.225 & 3.968 & 4.472 & 0.792 & 0.974 & 1.651 & 0.824 \\
		& SB-NCSN++(1) & 25.16 & 0.0155 & 3.387 & 4.087 & 3.148 & 3.665 & 3.892 & 0.787 & 0.962 & 1.669 & 0.791 \\
		& SBCTM & 65.98 & 0.02 & 3.523 & 4.128 & 3.298 & 3.906 & 4.574 & 0.870 & 0.986 & 2.835 & 0.898 \\
		& SB-UFOGen & 25.16 & 0.0152 & 3.443 & 4.007 & 3.161 & 3.76 & 3.719 & 0.793 & 0.964 & 1.754 & 0.803 \\
        & ZipEnhancer & 2.04 & 0.105 & 3.578 & \textbf{\underline{4.136}} & \textbf{\underline{3.35}} & 4.015 & 4.289 & \textbf{\underline{0.922}} & \textbf{\underline{0.992}} & \textbf{\underline{3.419}} & \textbf{\underline{0.952}} \\
		& Mamba-base & 3.61 & \textbf{\underline{0.0048}} & 3.515 & 4.053 & 3.253 & 3.997 & 4.468 & 0.887 & 0.99 & 2.73 & 0.916 \\
        & FM-Mamba & 3.93 & \textbf{\underline{0.0048}} & 3.446 & 4.057 & 3.192 & 3.783 & 4.088 & 0.802 & 0.968 & 1.964 & 0.843 \\
		& SBM & 3.93 & \textbf{\underline{0.0048}} & \textbf{\underline{3.751}} & 4.129 & 3.292 & \textbf{\underline{4.187}} & \textbf{\underline{4.657}} & 0.893 & 0.99 & 2.825 & 0.921 \\ 
		\midrule
		\multirow{9}{1.3cm}{VoiceBank-Demand}
		& SB-NCSN++(50) & 25.16 & 0.767 & 3.415 & 4.005 & 3.134 & 3.55 & 4.237 & 0.764 & 0.942 & 1.887 & 0.753 \\
		& SB-NCSN++(10) & 25.16 & 0.156 & 3.382 & 4.05 & 3.131 & 3.61 & 4.688 & 0.795 & 0.953 & 2.112 & 0.79 \\
		& SB-NCSN++(1) & 25.16 & 0.0155 & 3.337 & 4.087 & 3.106 & 3.394 & 4.41 & 0.778 & 0.934 & 2.165 & 0.76 \\
		& SBCTM & 65.98 & 0.02 & 3.4 & 4.085 & 3.156 & 3.533 & 4.646 & 0.891 & 0.978 & 3.558 & 0.868 \\
		& SB-UFOGen & 25.16 & 0.0152 & 3.384 & 4.06 & 3.133 & 3.371 & 4.132 & 0.768 & 0.926 & 2.152 & 0.758 \\
        & ZipEnhancer & 2.04 & 0.105 & \textbf{\underline{3.484}} & 3.988 & 3.183 & 3.544 & 4.621 & \textbf{\underline{0.925}} & \textbf{\underline{0.986}} & \textbf{\underline{3.628}} & \textbf{\underline{0.894}} \\
		& Mamba-base & 3.61 & \textbf{\underline{0.0048}} & 3.294 & 3.904 & 2.969 & 3.498 & 4.441 & 0.814 & 0.975 & 2.612 & 0.831 \\
        & FM-Mamba & 3.93 & \textbf{\underline{0.0048}} & 3.269 & 3.964 & 2.982 & 3.373 & 4.267 & 0.761 & 0.951 & 2.267 & 0.771 \\
		& SBM & 3.93 & \textbf{\underline{0.0048}} & 3.412 & \textbf{\underline{4.135}} & \textbf{\underline{3.253}} & \textbf{\underline{3.616}} & \textbf{\underline{4.737}} & 0.820 & 0.981 & 3.503 & 0.891 \\
		\bottomrule
	\end{tabular}
	\label{tab:test1}
\end{table*}

\begin{table}[t]
	\centering
	\caption{Results on DNS Real Recordings. Intrusive metrics are not included due to the absence of clean ground truth.}
    \tablescriptsize
	\renewcommand\arraystretch{0.75}
	\begin{tabular}{p{0.8cm} l c c c c c}
		\toprule
		Test set & Model & SIG$\uparrow$ & BAK$\uparrow$ & OVRL$\uparrow$ & \makecell[c]{P808\\MOS$\uparrow$} & NISQA$\uparrow$ \\
		\midrule
		\multirow{9}{0.8cm}{DNS Real Recordings} 
		& SB-NCSN++(50) & 3.369 & 3.939 & 3.052 & 3.685 & 3.604 \\
		& SB-NCSN++(10) & 3.332 & 3.933 & 3.024 & 3.723 & 3.865 \\
		& SB-NCSN++(1)  & 3.297 & 3.924 & 2.979 & 3.514 & 3.541 \\
		& SBCTM & 3.224 & 3.841 & 2.894 & 3.506 & 3.647 \\
		& SB-UFOGen & 3.255 & 3.838 & 2.911 & 3.485 & 3.196 \\
        & ZipEnhancer & 3.323 & 3.918 & 3.002 & 3.639 & 3.264 \\
		& Mamba-base & 3.284 & 3.974 & 2.993 & 3.683 & 3.644 \\
        & FM-Mamba & 3.32 & 4.031 & 3.052 & 3.562 & 3.797 \\
		& SBM & \textbf{\underline{3.459}} & \textbf{\underline{4.106}} & \textbf{\underline{3.198}} & \textbf{\underline{3.742}} & \textbf{\underline{3.937}} \\
		\bottomrule
	\end{tabular}
	\label{tab:test2}
\end{table}


\section{Experiments and Results}
\label{sec:exp}

\subsection{Experimental Setup}
\label{sec:imple}
To cover common real-life scenarios, we focus on joint denoising and dereverberation tasks for speech enhancement.
Audio clips are arbitrarily segmented into 3-second segments on-the-fly during training.  
For the STFT representation, we use a NFFT of 512, a Hann window of length 480 and a hop length of 160.
The SBM model is trained using AdamW optimizer with a cosine learning rate scheduler starting from $lr=1e-3$.
The hyper-parameters of the loss $\lambda_1, \lambda_2, \lambda_3, \lambda_4$ are set to 1000, 1000, 500, 500.

\subsection{Datasets and Metrics}
The training datasets used in our experiments include clean speeches, noises and room impulse responses (RIRs). Clean speeches are collected from DNS Challenge clean data~\cite{reddy2020interspeech}, AI-Shell3~\cite{shi2020aishell} and LibriSpeech~\cite{panayotov2015librispeech}, totaling around 800 hours. Noises are collected from FSD50K~\cite{fonseca2021fsd50k}, DNS Challenge~\cite{reddy2020interspeech} and self-collected noises from various environments. RIRs are collected from SLR26/28~\cite{ko2017study} and simulated using pyroomacoustics~\cite{scheibler2018pyroomacoustics}. 
The paired degraded speeches are simulated using the above clean speeches, noises and RIRs with a signal-to-noise ratio (SNR) in $[-10, 20]$ that can cover a wide range of daily-life scenarios. All audio data is pre-processed as \qty{16}{kHz}.   

We evaluate SBM on four benchmark test sets: DNS Challenge Real Recordings, Synthetic With Reverb, Synthetic Without Reverb~\cite{reddy2020interspeech} and VoiceBank-Demand test set~\cite{botinhao2016investigating}.

We adopt a set of metrics to evaluate speech enhancement:
DNSMOS~\cite{reddy2021dnsmos} (\textit{SIG} for signal quality, \textit{BAK} for noise quality, \textit{OVRL} and \textit{P808MOS} for overall quality), \textit{NISQA}~\cite{mittag2021nisqa} for perceptual quality, \textit{SpeechBERTScore}~\cite{saeki2024speechbertscore} for semantic similarity, \textit{Similarity}~\cite{zhang2025anyenhance} for speaker cosine similarity, 
and traditional alignment-sensitive metrics (\textit{PESQ}~\cite{beerends2005extension}, \textit{ESTOI}~\cite{jensen2016algorithm}).
We use the same pretrained model as \cite{zhang2025anyenhance} for \textit{SpeechBERTScore} and \textit{Similarity}.
Although employing a data prediction loss, SBM reconstructs speech details via trajectory guidance rather than yielding statistical averages~\cite{richter2023speech}. Since strictly alignment-sensitive metrics often penalize deviated structures more heavily than over-smoothed predictions~\cite{blau2018perception}, alignment-agnostic metrics (\textit{SIG}, \textit{BAK}, \textit{OVRL}, \textit{P808MOS}, \textit{NISQA}, \textit{SpeechBERTScore}, \textit{Similarity}) serve as our primary indicators of model performance, alongside alignment-sensitive metrics for a comprehensive assessment. 
 
\subsection{Baseline Models}  
We validate SBM by comparing it with a group of baselines:

\textbf{SB-NCSN++}~\cite{jukic2024schr}: The classical model of SB for SE. We include three versions for 50-steps inference (\textbf{SB-NCSN++(50)}), 10-steps inference (\textbf{SB-NCSN++(10)}) and 1-step inference (\textbf{SB-NCSN++(1)}). 
The 1-step version follows the same inference strategy as SBM.
Due to the absence of public checkpoints, we re-trained models following the paper and open-sourced codes using the same datasets for SBM.
We also tried different losses following the design of SBM and literature~\cite{jukic2024schr,li2025bridge,kong2025a2sb}, and report SB-NCSN++ baselines with the best results. 

\textbf{SBCTM}~\cite{nishigori2025schrodinger}: Built on top of \cite{jukic2024schr,richter2025investigating}, SBCTM involves consistency trajectory modeling~\cite{kim2023consistency} to achieve 1-step inference.
We used their open-sourced checkpoint in our experiments.

\textbf{SB-UFOGen}~\cite{han2025few}: Built on top of \cite{jukic2024schr}, SB-UFOGen incorporates adversarial training~\cite{xu2024ufogen} to achieve 1-step inference.
Due to the absence of open-sourced checkpoints, we re-trained this baseline as we did for SB-NCSN++.

\textbf{ZipEnhancer}~\cite{wang2025zipenhancer}: A state-of-the-art discriminative model. They published checkpoint for the DNS dataset~\cite{ZipEnhancerCkpt}. For the VoiceBank-Demand test set, lacking a public checkpoint, we computed metrics using the officially released inference results~\cite{ZipEnhancerVBdemand}. 


\textbf{Mamba-base}: To justify the SB training paradigm, this baseline refers to a predictive-mapping trained Mamba-based model. For a fair comparison, we trained this baseline using the same datasets and the same Mamba-based backbone as in SBM.

\textbf{FM-Mamba}: To align with recent generative modeling trends and strengthen our evaluation, we introduce a flow matching (FM)~\cite{lipman2022flow} variant. This baseline shares the exact Mamba backbone and one-step inference protocol with SBM, but computes the intermediate states $x_t$ via FM instead of the SB formulation. Specifically, we computed the states $x_t$ using the OT-CFM formulation~\cite{korostik2025modifying,wang2026rethinking}. This baseline aims to validate whether the SB objective yields superiority over the FM paradigm in our experiments.

\subsection{Experimental Results and Discussion}
As shown in Table~\ref{tab:test1} and Table~\ref{tab:test2}, 
SBM demonstrates strong generalization in complex acoustic environments. 
On the DNS Real Recordings that represent in-the-wild degraded audio, SBM achieves the highest scores across all metrics. 
Similarly, on the DNS With Reverb, SBM leads on \textit{SIG}, \textit{BAK}, \textit{OVRL}, \textit{P808MOS}, \textit{SpeechBERTScore}, \textit{Similarity}, \textit{PESQ} and \textit{ESTOI}.
These results indicate the effectiveness of SBM in jointly addressing real-world noise and reverberation artifacts.

On the DNS No Reverb and VoiceBank-Demand test sets that only include noise artifacts, 
SBM performs competitively against the discriminative model ZipEnhancer.
While the results of ZipEnhancer demonstrate the efficacy of its multi-objective training (e.g., via PESQ-based GAN and phase losses)~\cite{wang2025zipenhancer}, SBM achieves on-par performance via trajectory-guided reconstruction.
Furthermore, by incorporating 25\% double-talk scenarios during training to preserve secondary speakers, SBM diverges from test cases requiring single-speaker isolation. Consequently, this discrepancy penalizes SBM on certain metrics like \textit{SpeechBERTScore}, which reflects a strategic design choice for realistic multi-speaker environments rather than a deficit in enhancement capability.

According to Table~\ref{tab:test1}, SBM achieves the lowest real-time factor (RTF). Beyond computational efficiency, it maintains a small algorithmic latency (see Sec~\ref{sec:methodsub2}), suggesting its potential for practical low-latency applications.  



\begin{figure}[t!]
	\centering 
	\includegraphics[width=\columnwidth]{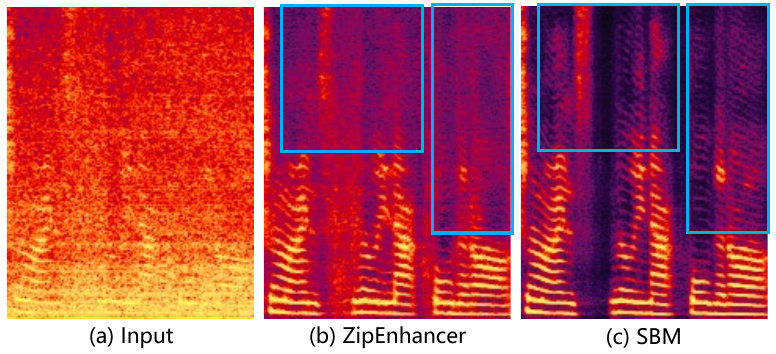}
	\caption{Spectrogram comparison in a challenging real-world case. SBM successfully reconstructs frequency harmonics, reflecting learned structural priors.}
	\label{fig:generativeNature}
\end{figure}

We further examine the impact of the training paradigm of SBM.
Although employing a deterministic one-step inference akin to conventional prediction models~\cite{wang2026rethinking}, SBM receives principled generative guidance by training using the SB OT trajectory. 
As shown in Figure~\ref{fig:generativeNature}, under severe real-world degradation, SBM successfully reconstructs mid-to-high frequency harmonics, whereas the discriminative baseline exhibits typical over-smoothing. 
This implies that SBM learns the structural priors, enabling the reconstruction of fine-grained harmonics rather than defaulting to a statistical average.  

The empirical results also justify the advantages of the Mamba architecture and the SB training paradigm. 
SB-NCSN++, SBCTM and SB-UFOGen are all built based on the NCSN++ backbone that is most commonly applied in SB for SE models. 
While CTM and adversarial training techniques accelerate inference while maintaining model performance~\cite{han2025few,nishigori2025schrodinger}, they fail in addressing the size and efficiency problems intrinsic to the NCSN++ structure. 
In comparison, SBM brings significant improvements, showing the efficacy of model structural change under the scope of SB training paradigm.
On the other hand, by altering the training paradigm, SBM outperforms Mamba-base and FM-Mamba without RTF degradation. Although the incorporation of timestep embedding slightly enlarges the SBM model, this barely influences inference efficiency. The same backbone structure also implies that edge-side optimization techniques for Mamba are also applicable to SBM, further facilitating the deployment of SBM.    


\subsection{Ablation Studies}
To further evaluate the synergy between backbones and training paradigms, we conducted ablation studies comparing Mamba against Multi-Head Self-Attention (MHSA) and Long Short-Term Memory (LSTM).
We replaced all Mamba blocks in Figure~\ref{fig:overview} with MHSA or LSTM, and trained models using the SB paradigm and the mapping paradigm respectively. 
To provide each backbone with a comparable capacity for a fair comparison, 
the design, hidden dimensions and hyper-parameters of MHSA and LSTM blocks are carefully configured based on their structural correspondence to Mamba~\cite{gu2023mamba,gu2022efficiently,dao2024transformers}, leading to the different model sizes shown in Table~\ref{tab:ablation}. Table~\ref{tab:ablation} shows the comparison results on DNS Real Recordings.

The results show that the SB paradigm consistently leads across all architectures, demonstrating the effectiveness of training on intermediate states $x_t$ as a form of trajectory-based supervision. 
Unlike static point-to-point mapping, SB leverages the optimal transport path to provide a continuous-time view of the speech enhancement evolution process.
Notably, Mamba achieves superior performance than MHSA and LSTM, indicating that its selective state-space mechanism could be more suitable for internalizing the evolution dynamics. 
While future work will further investigate the contribution of the trajectory modeling-based training, these results highlight the effective synergy between the SB paradigm and the Mamba architecture.


\begin{table}[t]
	\centering
	\caption{Ablation study results on DNS Real Recordings. \textit{(M)} stands for the mapping training paradigm and \textit{(SB)} stands for the SB training paradigm.}
    \tablescriptsize
	\renewcommand\arraystretch{0.75}
	\begin{tabular}{l c c c c c c}
		\toprule
		Model & Parameters & SIG$\uparrow$ & BAK$\uparrow$ & OVRL$\uparrow$ & \makecell[c]{P808\\MOS$\uparrow$} & NISQA$\uparrow$ \\
		\midrule
        MHSA\textit{(M)} & 8.37M & 2.975 & 3.809 & 2.676 & 3.233 & 2.358 \\
		MHSA\textit{(SB)} & 8.41M & \textbf{3.221} & \textbf{3.941} & \textbf{2.927} & \textbf{3.60} & \textbf{3.24} \\
        \midrule
		LSTM\textit{(M)} & 17.38M & 3.083 & 3.904 & 2.799 & 3.331 & 2.585 \\
		LSTM\textit{(SB)} & 17.42M & \textbf{3.289} & \textbf{3.965} & \textbf{3.00} & \textbf{3.674} & \textbf{3.819} \\
        \midrule
		Mamba\textit{(M)} & 3.61M & 3.284 & 3.974 & 2.993 & 3.683 & 3.644 \\
        Mamba\textit{(SB)} & 3.93M & \textbf{\underline{3.459}} & \textbf{\underline{4.106}} & \textbf{\underline{3.198}} & \textbf{\underline{3.742}} & \textbf{\underline{3.937}} \\
		\bottomrule
	\end{tabular}
	\label{tab:ablation}
\end{table}

\section{Conclusion}
\label{sec:con}
In this paper, we introduced Schrödinger Bridge Mamba (SBM), a novel one-step speech enhancement model guided by the generative Schrödinger Bridge (SB) paradigm and the Mamba architecture. Extensive evaluations demonstrate that SBM delivers state-of-the-art performance in complex real-world and reverberant scenarios, while maintaining a highly competitive real-time factor and a lightweight model size. 
Furthermore, ablation studies suggest that this robust performance likely benefits from the structural alignment between SB's optimal transport trajectory and Mamba's continuous-time dynamics.
Moving forward, we aim to extend SBM to broader audio processing tasks like super-resolution and semantic-level restoration, and further investigate its underlying mechanism, thereby offering valuable insights for achieving high-fidelity speech processing without compromising the efficiency required by real-world applications. 



\newpage

\section{Generative AI Use Disclosure}
In compliance with ISCA policy, the authors declare that generative AI tools were utilized solely for editing and polishing the language of this manuscript to improve readability. No generative AI was used to generate core ideas, design experiments, write any major part, or produce data analyses. The authors take full responsibility for the content and integrity of this work.

\bibliographystyle{IEEEtran}
\bibliography{mybib}

@article{richter2023speech,
	title={{Speech enhancement and dereverberation with diffusion-based generative models}},
	author={Richter, Julius and Welker, Simon and Lemercier, Jean-Marie and Lay, Bunlong and Gerkmann, Timo},
	journal={IEEE/ACM Transactions on Audio, Speech, and Language Processing},
	volume={31},
	pages={2351--2364},
	year={2023},
	publisher={IEEE}
}

@inproceedings{jukic2024schr,
  title={{Schr{\"o}dinger Bridge for Generative Speech Enhancement}},
  author={Juki{\'c}, Ante and Korostik, Roman and Balam, Jagadeesh and Ginsburg, Boris},
  booktitle={Proc. Interspeech 2024},
  pages={1175--1179},
  _year={2024}
}

@article{wang2024diffusion,
	title={{Diffusion-based Speech Enhancement with Schr{\"o}dinger Bridge and Symmetric Noise Schedule}},
	author={Wang, Siyi and Liu, Siyi and Harper, Andrew and Kendrick, Paul and Salzmann, Mathieu and Cernak, Milos},
	journal={arXiv preprint arXiv:2409.05116},
	year={2024}
}

@inproceedings{richter2025investigating,
	title={{Investigating training objectives for generative speech enhancement}},
	author={Richter, Julius and De Oliveira, Danilo and Gerkmann, Timo},
	booktitle={IEEE International Conference on Acoustics, Speech and Signal Processing (ICASSP)},
	pages={1--5},
	year={2025},
	organization={IEEE}
}

@inproceedings{han2025few,
  title={{Few-step Adversarial Schr{\"o}dinger Bridge for Generative Speech Enhancement}},
  author={Han, Seungu and Lee, Sungho and Lee, Juheon and Lee, Kyogu},
  booktitle={Proceedings of the Annual Conference of the International Speech Communication Association, INTERSPEECH},
  pages={2380--2384},
  year={2025},
  organization={International Speech Communication Association}
}

@inproceedings{xu2024ufogen,
	title={{UFOGen: You forward once large scale text-to-image generation via diffusion GANs}},
	author={Xu, Yanwu and Zhao, Yang and Xiao, Zhisheng and Hou, Tingbo},
	booktitle={Proceedings of the IEEE/CVF Conference on Computer Vision and Pattern Recognition},
	pages={8196--8206},
	year={2024}
}

@inproceedings{nishigori2025schrodinger,
  title={{Schr{\"o}dinger bridge consistency trajectory models for speech enhancement}},
  author={Nishigori, Shuichiro and Saito, Koichi and Murata, Naoki and Hirano, Masato and Takahashi, Shusuke and Mitsufuji, Yuki},
  booktitle={2025 IEEE Workshop on Applications of Signal Processing to Audio and Acoustics (WASPAA)},
  pages={1--5},
  _year={2025},
  _organization={IEEE}
}

@inproceedings{kim2023consistency,
  title={{Consistency Trajectory Models: Learning Probability Flow ODE Trajectory of Diffusion}},
  author={Kim, Dongjun and Lai, Chieh-Hsin and Liao, Wei-Hsiang and Murata, Naoki and Takida, Yuhta and Uesaka, Toshimitsu and He, Yutong and Mitsufuji, Yuki and Ermon, Stefano},
  booktitle={The Twelfth International Conference on Learning Representations}
}

@inproceedings{li2025bridge,
	title={{Bridge-SR: Schr{\"o}dinger bridge for efficient SR}},
	author={Li, Chang and Chen, Zehua and Bao, Fan and Zhu, Jun},
	booktitle={IEEE International Conference on Acoustics, Speech and Signal Processing (ICASSP)},
	pages={1--5},
	year={2025},
	_organization={IEEE}
}

@article{kong2025a2sb,
	title={{A2SB: Audio-to-Audio Schrodinger Bridges}},
	author={Kong, Zhifeng and Shih, Kevin J and Nie, Weili and Vahdat, Arash and Lee, Sang-gil and Santos, Joao Felipe and Jukic, Ante and Valle, Rafael and Catanzaro, Bryan},
	journal={arXiv preprint arXiv:2501.11311},
	year={2025}
}

@inproceedings{liu20232,
  title={{I2SB: image-to-image Schr{\"o}dinger bridge}},
  author={Liu, Guan-Horng and Vahdat, Arash and Huang, De-An and Theodorou, Evangelos A and Nie, Weili and Anandkumar, Anima},
  booktitle={Proceedings of the 40th International Conference on Machine Learning},
  pages={22042--22062},
  year={2023}
}

@article{chen2023schrodinger,
	title={{Schrodinger bridges beat diffusion models on text-to-speech synthesis}},
	author={Chen, Zehua and He, Guande and Zheng, Kaiwen and Tan, Xu and Zhu, Jun},
	journal={arXiv preprint arXiv:2312.03491},
	year={2023}
}

@inproceedings{songscore,
	title={{Score-Based Generative Modeling through Stochastic Differential Equations}},
	author={Song, Yang and Sohl-Dickstein, Jascha and Kingma, Diederik P and Kumar, Abhishek and Ermon, Stefano and Poole, Ben},
	booktitle={International Conference on Learning Representations}
}

@article{vahdat2021score,
	title={{Score-based generative modeling in latent space}},
	author={Vahdat, Arash and Kreis, Karsten and Kautz, Jan},
	journal={Advances in neural information processing systems},
	volume={34},
	pages={11287--11302},
	year={2021}
}

@inproceedings{gu2023mamba,
  title={{Mamba: Linear-time sequence modeling with selective state spaces}},
  author={Gu, Albert and Dao, Tri},
  booktitle={First conference on language modeling}
}

@article{quan2024multichannel,
	title={{Multichannel long-term streaming neural speech enhancement for static and moving speakers}},
	author={Quan, Changsheng and Li, Xiaofei},
	journal={IEEE Signal Processing Letters},
	volume={31},
	pages={2295--2299},
	year={2024},
	publisher={IEEE}
}

@inproceedings{chao2024investigation,
	title={{An investigation of incorporating mamba for speech enhancement}},
	author={Chao, Rong and Cheng, Wen-Huang and La Quatra, Moreno and Siniscalchi, Sabato Marco and Yang, Chao-Han Huck and Fu, Szu-Wei and Tsao, Yu},
	booktitle={IEEE Spoken Language Technology Workshop (SLT)},
	pages={302--308},
	year={2024},
	_organization={IEEE}
}

@inproceedings{chao2025universal,
  title={{Universal Speech Enhancement with Regression and Generative Mamba}},
  author={Chao, Rong and Nasretdinov, Rauf and Wang, Yu-Chiang Frank and Jukic, Ante and Fu, Szu-Wei and Tsao, Yu},
  booktitle={Proc. Interspeech 2025},
  pages={888--892},
  _year={2025}
}

@inproceedings{hu2024zigma,
  title={{Zigma: A DiT-style zigzag mamba diffusion model}},
  author={Hu, Vincent Tao and Baumann, Stefan Andreas and Gui, Ming and Grebenkova, Olga and Ma, Pingchuan and Fischer, Johannes and Ommer, Bj{\"o}rn},
  booktitle={European conference on computer vision},
  pages={148--166},
  year={2024},
  organization={Springer}
}

@inproceedings{wang2025zipenhancer,
  title={{ZipEnhancer: dual-path down-up sampling-based zipformer for monaural speech enhancement}},
  author={Wang, Haoxu and Tian, Biao},
  booktitle={IEEE International Conference on Acoustics, Speech and Signal Processing (ICASSP)},
  pages={1--5},
  year={2025},
  _organization={IEEE}
}

@article{chen2021stochastic,
	title={{Stochastic control liaisons: Richard sinkhorn meets gaspard monge on a Schrodinger Bridge}},
	author={Chen, Yongxin and Georgiou, Tryphon T and Pavon, Michele},
	journal={Siam Review},
	volume={63},
	number={2},
	pages={249--313},
	year={2021},
	publisher={SIAM}
}

@inproceedings{schrodinger1932theorie,
	title={{Sur la th{\'e}orie relativiste de l'{\'e}lectron et l'interpr{\'e}tation de la m{\'e}canique quantique}},
	author={Schr{\"o}dinger, Erwin},
	booktitle={Annales de l'institut Henri Poincar{\'e}},
	volume={2},
	number={4},
	pages={269--310},
	year={1932}
}

@article{kalman1960new,
	title={{A new approach to linear filtering and prediction problems}},
	author={Kalman, Rudolph Emil},
	year={1960}
}

@article{quan2024spatialnet,
	title={{SpatialNet: Extensively learning spatial information for multichannel joint speech separation, denoising and dereverberation}},
	author={Quan, Changsheng and Li, Xiaofei},
	journal={IEEE/ACM Transactions on Audio, Speech, and Language Processing},
	volume={32},
	pages={1310--1323},
	year={2024},
	publisher={IEEE}
}

@article{reddy2020interspeech,
  title={{The INTERSPEECH 2020 Deep Noise Suppression Challenge: Datasets, Subjective Testing Framework, and Challenge Results}},
  author={Reddy, Chandan KA and Gopal, Vishak and Cutler, Ross and Beyrami, Ebrahim and Cheng, Roger and Dubey, Harishchandra and Matusevych, Sergiy and Aichner, Robert and Aazami, Ashkan and Braun, Sebastian and others},
  journal={Interspeech 2020},
  _year={2020},
  publisher={ISCA}
}

@inproceedings{shi2020aishell,
  title={{AISHELL-3: A Multi-Speaker Mandarin TTS Corpus}},
  author={Shi, Yao and Bu, Hui and Xu, Xin and Zhang, Shaoji and Li, Ming},
  booktitle={Proc. Interspeech 2021},
  pages={2756--2760},
  year={2021}
}

@inproceedings{panayotov2015librispeech,
	title={{Librispeech: an ASR corpus based on public domain audio books}},
	author={Panayotov, Vassil and Chen, Guoguo and Povey, Daniel and Khudanpur, Sanjeev},
	booktitle={IEEE international conference on acoustics, speech and signal processing (ICASSP)},
	pages={5206--5210},
	year={2015},
	_organization={IEEE}
}

@article{fonseca2021fsd50k,
	title={{FSD50k: an open dataset of human-labeled sound events}},
	author={Fonseca, Eduardo and Favory, Xavier and Pons, Jordi and Font, Frederic and Serra, Xavier},
	journal={IEEE/ACM Transactions on Audio, Speech, and Language Processing},
	volume={30},
	pages={829--852},
	year={2021},
	publisher={IEEE}
}

@inproceedings{ko2017study,
	title={{A study on data augmentation of reverberant speech for robust speech recognition}},
	author={Ko, Tom and Peddinti, Vijayaditya and Povey, Daniel and Seltzer, Michael L and Khudanpur, Sanjeev},
	booktitle={IEEE international conference on acoustics, speech and signal processing (ICASSP)},
	pages={5220--5224},
	year={2017},
	_organization={IEEE}
}

@inproceedings{scheibler2018pyroomacoustics,
	title={{Pyroomacoustics: A python package for audio room simulation and array processing algorithms}},
	author={Scheibler, Robin and Bezzam, Eric and Dokmani{\'c}, Ivan},
	booktitle={IEEE international conference on acoustics, speech and signal processing (ICASSP)},
	pages={351--355},
	year={2018},
	_organization={IEEE}
}

@inproceedings{botinhao2016investigating,
	title={{Investigating RNN-based speech enhancement methods for noise-robust text-to-speech}},
	author={Botinhao, Cassia Valentini and Wang, Xin and Takaki, Shinji and Yamagishi, Junichi},
	booktitle={9th ISCA speech synthesis workshop},
	pages={159--165},
	year={2016}
}

@inproceedings{reddy2021dnsmos,
	title={{DNSMOS: A non-intrusive perceptual objective speech quality metric to evaluate noise suppressors}},
	author={Reddy, Chandan KA and Gopal, Vishak and Cutler, Ross},
	booktitle={IEEE International Conference on Acoustics, Speech and Signal Processing (ICASSP)},
	pages={6493--6497},
	year={2021},
	_organization={IEEE}
}

@inproceedings{mittag2021nisqa,
  title={{NISQA: A Deep CNN-Self-Attention Model for Multidimensional Speech Quality Prediction with Crowdsourced Datasets}},
  author={Mittag, Gabriel and Naderi, Babak and Chehadi, Assmaa and M{\"o}ller, Sebastian},
  booktitle={Proc. Interspeech 2021},
  pages={2127--2131},
  _year={2021}
}

@inproceedings{saeki2024speechbertscore,
  title={{SpeechBERTScore: Reference-Aware Automatic Evaluation of Speech Generation Leveraging NLP Evaluation Metrics}},
  author={Saeki, Takaaki and Maiti, Soumi and Takamichi, Shinnosuke and Watanabe, Shinji and Saruwatari, Hiroshi},
  booktitle={Proceedings of the Annual Conference of the International Speech Communication Association, INTERSPEECH},
  pages={4943--4947},
  year={2024},
  organization={International Speech Communication Association}
}

@article{zhang2025anyenhance,
  title={{Anyenhance: A unified generative model with prompt-guidance and self-critic for voice enhancement}},
  author={Zhang, Junan and Yang, Jing and Fang, Zihao and Wang, Yuancheng and Zhang, Zehua and Wang, Zhuo and Fan, Fan and Wu, Zhizheng},
  journal={IEEE Transactions on Audio, Speech and Language Processing},
  year={2025},
  _publisher={IEEE}
}

@article{beerends2005extension,
	title={{Extension of ITU-T recommendation P. 862 PESQ towards measuring speech intelligibility with vocoders}},
	author={Beerends, John G and Van Wijngaarden, Sander and Van Buuren, Ronald},
	year={2005}
}

@article{jensen2016algorithm,
	title={{An algorithm for predicting the intelligibility of speech masked by modulated noise maskers}},
	author={Jensen, Jesper and Taal, Cees H},
	journal={IEEE/ACM Transactions on Audio, Speech, and Language Processing},
	volume={24},
	number={11},
	pages={2009--2022},
	year={2016},
	publisher={IEEE}
}

@misc{ZipEnhancerCkpt,
  title = {{ZipEnhancer checkpoint for DNS dataset}},
  howpublished = "\url{https://modelscope.cn/models/iic/speech_zipenhancer_ans_multiloss_16k_base}",
  year = {2025}, 
  note = "[Online; accessed 19-January-2026]"
}

@misc{ZipEnhancerVBdemand,
  title = {{ZipEnhancer-enhanced VoiceBank-Demand test set}},
  howpublished = "\url{https://github.com/ZipEnhancer/ZipEnhancer/tree/main/enhanced_wavs/VoiceBank%2BDEMAND}",
  year = {2024}, 
  note = "[Online; accessed 19-January-2026]"
}

@inproceedings{lipman2022flow,
  title={{Flow matching for generative modeling}},
  author={Lipman, Yaron and Chen, Ricky TQ and Ben-Hamu, Heli and Nickel, Maximilian and Le, Matt},
  booktitle={The Eleventh International Conference on Learning Representations},
  year={2022}
}

@inproceedings{korostik2025modifying,
  title={{Modifying flow matching for generative speech enhancement}},
  author={Korostik, Roman and Nasretdinov, Rauf and Juki{\'c}, Ante},
  booktitle={IEEE International Conference on Acoustics, Speech and Signal Processing (ICASSP)},
  pages={1-5},
  year={2025},
  _organization={IEEE}
}

@inproceedings{bondaschi2025markov,
  title={{From Markov to Laplace: How Mamba In-Context Learns Markov Chains}},
  author={Bondaschi, Marco and Rajaraman, Nived and Wei, Xiuying and Ramchandran, Kannan and Pascanu, Razvan and Gulcehre, Caglar and Gastpar, Michael and Makkuva, Ashok Vardhan},
  booktitle={ICLR 2025 Workshop: XAI4Science: From Understanding Model Behavior to Discovering New Scientific Knowledge}
}

@article{wang2026rethinking,
  title={{Rethinking Flow and Diffusion Bridge Models for Speech Enhancement}},
  author={Wang, Dahan and Gao, Jun and Lei, Tong and Hu, Yuxiang and Zhu, Changbao and Chen, Kai and Lu, Jing},
  journal={arXiv preprint arXiv:2602.18355},
  year={2026}
}

@inproceedings{gu2022efficiently,
  title={{Efficiently Modeling Long Sequences with Structured State Spaces}},
  author={Gu, Albert and Goel, Karan and R\'e, Christopher},
  booktitle={The International Conference on Learning Representations ({ICLR})},
  year={2022}
}

@inproceedings{dao2024transformers,
  title={{Transformers are SSMs: generalized models and efficient algorithms through structured state space duality}},
  author={Dao, Tri and Gu, Albert},
  booktitle={Proceedings of the 41st International Conference on Machine Learning},
  pages={10041--10071},
  year={2024}
}

@inproceedings{blau2018perception,
  title={{The perception-distortion tradeoff}},
  author={Blau, Yochai and Michaeli, Tomer},
  booktitle={Proceedings of the IEEE conference on computer vision and pattern recognition},
  pages={6228--6237},
  year={2018}
}

\end{document}